\documentstyle[amssymb,times,pramana,epsf,floats]{ias}  
\def\nue{{\nu_e}}

\def\lsim{\raise0.3ex\hbox{$\;<$\kern-0.75em\raise-1.1ex\hbox{$\sim\;$}}} 
\newcommand{\be}{\begin{eqnarray*}}
\newcommand{\ee}{\end{eqnarray*}}
\newcommand{\beq}{\begin{equation}}
\newcommand{\eeq}{\end{equation}}

\newcommand{\bref}{\begin{flushright}\tiny\Magenta}
\newcommand{\eref}{\end{flushright}}

\begin{document}

\title
{ Working Group Report: 
Neutrino and Astroparticle Physics}

\author{
{\it Coordinators}:{\large{ Srubabati Goswami and Raghavan Rangarajan}}  
\\
{\it Working Group Members}: 
K. Agashe, A. Bandyopadhyay,
K. Bhattacharya,
B. Brahmachari, 
C. Burgess, E.J. Chun,
D. Choudhury,
P.K.Das, A. Dighe,
R. Godbole,
S.Goswami,
N. Gupta,
M. Kaplinghat,
D. Indumathi,
J. Forshaw,
Y.Y. Keum,
B. Layek,
D. Majumdar,
N. Mahajan,
P. Mehta,
R.N. Mohapatra,
N. Mondal,
S. More,
Y. Nir,
S. Pakvasa,
M.K. Parida,
M. Ravikumar,
G. Rajasekaran,
P. Ramadevi,
R. Rangarajan,
S.D. Rindani,
D.P. Roy,
P. Roy ,
N. Sahu,
A. Samanta,
Y. Shadmi,
A.M. Srivastava,
S. Uma Sankar,
R. Vaidya,
U. Yajnik
}
\abstract{
This is the report of neutrino and astroparticle physics working group 
at WHEPP-8. 
We present the discussions carried out during the workshop 
on selected topics in the above fields and also indicate 
progress made subsequently.
The neutrino physics subgroup studied the possibilites of constraining 
neutrino masses, mixing and CPT violation in lepton sector
from future experiments. Neutrino mass models in the context of abelian 
horizontal symmetries, warped extra dimensions and in presence of 
triplet Higgs were studied.  
Effect of threshold corrections on radiative magnification of 
mixing angles was investigated. 
The astroparticle physics subgroup focused on how various particle physics
inputs affect the CMBR fluctuation spectrum, and on brane cosmology.
This report also contains an introduction on how to use the publicly
available code CMBFAST to calculate the CMBR fluctuations.
}
\maketitle 

\section{Introduction} 
The working group on Neutrino and Astroparticle Physics had two main themes. 
Neutrinos and Cosmology. 

The Neutrino Physics subgroup 
studied the possibilites of  
determination of the neutrino oscillation parameters 
using almost pure $\nu_\mu$ beams in future superbeam experiments
as well as using pure $\nu_e$ beams in future $\beta$-beam experiments. 
CPT violation in neutrinos and possible outcomes of this 
in long baseline experiments and neutrino factories 
was posed as a possible  area of further study. 
Scenarios where each of the three neutrino flavours is contemplated 
to be a pseudo-Dirac state consisting of one active and one sterile 
partner and the possible effects of this on Supernova neutrino 
detection emerged as an intersting problem to be investigated 
further. 
Neutrino masses in the context of Abelian horizontal symmetries 
in the framework of Froggat-Nielsen mechanism were explored. 
Neutrino masses and mixings in warped extra dimensions 
were also looked into. The effect of threshold corrections 
on radiative magnifications and mixing for quasidegenarate 
neutrinos were studied. 
Triplet Higgs models for neutrino masses and possible tests of these
in upcoming collider experiments were discussed. Finally there 
were disucssions
on the possible areas that can be further explored in leptogenesis. 

The efforts of the Astroparticle Physics subgroup were focused on
understanding how fluctuations in the cosmic microwave background radiation are
affected by   various particle physics inputs
and on brane cosmology.  Calculating the effects of
particle physics inputs, such as neutrino properties, on the CMBR
requires
one to numerically solve the evolution of photons from decoupling to the
presentera.  Therefore,
this report contains a brief introduction on how to use CMBFAST, the publicly
available
code that allows one to calculate CMBR fluctuations as a function of
various input
parameters.  We then include a list of some particle physics related issues
that can have an
effect on the CMBR fluctuations and discuss how CMBFAST should be amended to
study these effects.  We further present the results of some prelimnary efforts to
modify
the code.  In brane cosmology, discussions focused on understanding inflation
and reheating in the context of brane universes, 
and on explaining the origin of dark energy in 
the context of supersymmetric ADD models.

Below
we disuss the work done in these two 
subgroups in two different sections. 

\section{Neutrinos} 
In the neutrino sector the two focal themes were  
{\it Precision Determination of Neutrino Oscillation Parameters} and 
{{\it Neutrino mass models}}. 
The three plenary talks in these topics were 
{{\it Long baseline neutrino experiments by S. Uma Sankar,  
Neutrino mass models and proton decay by R. N. Mohapatra, and 
India Based Neutrino Observatory by D. Indumathi}}
In addition the following seminars  
were arranged:  
\\ \\
$\bullet$ High Energy Astrophysical Neutrinos 
---{S. Pakvasa} \\
$\bullet$ A to Z symmetries of the Neutrino Mass Matrix 
---{G. Rajasekaran} \\
$\bullet$
CPT violation with Atmospheric Neutrinos
---{Poonam Mehta} \\
$\bullet$ Radiative magnification of
mixing for quasi degenerate neutrinos :
Effect of threshold corrections
---{M.K.Parida} \\
\\
Apart from this there were short presentations by the members of the 
Indian Neutrino Observatory (INO) collaboration in a session 
devoted to discussion on various issues related to the 
INO experiment. 
The working group formation and discussions of  topics were held 
on 8th January, 2004. 
Probir Roy led the disucussions on Neutrinos Mass models.  
The following topics were discussed and possibilites of further 
investigations explored: 
 \\ \\
$\bullet${PseudoDirac Neutrinos and their effect on supernova neutrino 
signal}\\
%A. Bandyopadhyay, A. Dighe, S. Goswami, 
%D. Majumdar, P. Mehta,
%S. Pakvasa.
$\bullet${Probing $\theta_{13}$ and sign of $\Delta_{31}$ with almost pure 
$\nu_\mu$ beams}\\
%A. Bandyopadhyay, A. Dighe, P.K. Das,
%S. Goswami, P. Mehta, S. Uma Sankar 
$\bullet$ {Studying Neutrino Oscillation parameters with pure $\nu_e$ beams}
\\
%A. Bandyopadhyay, S. Goswami, P. Mehta, S. Uma Sankar
$\bullet$ {Discussion on INO}\\ 
%{S. Chattopadhyaya,
%A. Dighe, D. Indumathi, S. Goswami, D. Majumdar, P. Mehta, N. Mondal,
%G. Rajasekaran, A. Raychaudhury, D.P. Roy, P. Roy, A. Samanta, 
%S. Uma Sankar}
$\bullet$ {CPT violation and possible effects in LBL experiments}\\
%{A. Dighe, P. Mehta}
%\item
$\bullet$ {Neutrino Masses in warped extra dimension}\\
%\\
%{K. Agashe, A. Dighe, P. Das, P. Mehta, P. Roy}
%\item
$\bullet$ 
{Supersymmetric Triplet Higgs Model for $M_{\nu}$: LE flavour violation
and HE collider test} \\
%\\
%{E.J. Chun, R. Godbole,Y. Keum, R.Vaidya,S.D. Rindani, J. Forshaw
%}
$\bullet$
{Leptogenesis} \\
$\bullet$ {Neutrino masses with Abelian Horizontal Symmetries}\\ 
%D. Indumathi, S. Goswami, Y. Shadmi, Y. Nir
%
%\item
$\bullet$
{Radiative Magnification of 
mixing for quasi degenerate neutrinos :
Effect of threshold corrections}
\\
%{R.N. Mohapatra, M.K.Parida, G. Rajasekaran}
%
%\end{enumerate}
%\newpage 

%\begin{enumerate}
%\item
\subsection{PseudoDirac Neutrinos and their effect on supernova neutrino 
signal \\
{\it A. Bandyopadhyay, D. Choudhury,  A. Dighe, S. Goswami, 
D. Majumdar, P. Mehta,
S. Pakvasa}}

We considered the scenario where each of the three active neutrinos  
may be considered to be a Pseudo-Dirac Neuttrino, 
composed of one active and one sterile 
state. 
Recently such a scenario has been considered in 
\cite{Beacom:2003eu}. 
The mass difference between each paired state is very small 
i.e they are almost degenerate and maximally mixed with 
$\Delta m^2 \sim 10^{-19}$ eV$^2$. 
In such a picture one will have 
additional  MSW resonances inside the supernova core.  
The mass difference of $\Delta m^2 \sim 10^{-19}$ eV$^2$ will also 
give rise to 
oscillation effects on the way from supernova to earth.   
The survival and conversion probabilities 
for this scenario will therefore be different from the standard 
three generation picture and 
the effect on the supernova neutrino signal may therefore be different. 
The propgation of the neutrinos inside the supernova for a such a scenario
was 
studied during the workshop. 
While this study was going on during the workshop
a similar paper came out in the spires 
\cite{Keranen:2004rg}. Efforts are on to do a more detailed study.  

%\item
{\subsection{Probing $\theta_{13}$ and sign of $\Delta_{31}$ with almost pure 
$\nu_\mu$ beams \\
A. Bandyopadhyay, A. Dighe, P.K. Das,
S. Goswami, P. Mehta, S. Uma Sankar} 
We propose to 
combine two $\nu$ superbeams at different baselines : \\
$\bullet$ 
$\nu$ superbeam from JHF with  L = 295 km and E = .75 GeV \\
$\bullet$
NuMI-Off Axis $\nu$ superbeam with L = 732 km and E = 1.5 
 GeV \\ 
The aim is to 
look for matter effects 
using two different baselines 
So one should first look 
at the appearance mode   
$
\nu_\mu \to \nu_e
$ and vary $\sin^2 2\theta_{13}$ between 0.1 to 0.25  
for which terms $\sim \Delta_{21}/\Delta_{31}$ can be neglected 
Then one can
determine the signal at J2K experiment for the above values of $\theta_{13}$ 
and check 
if the sign of $\Delta_{31}$ can be probed from NuMI Off-Axis experiment
for $\theta_{13}$ in the above range. 
The next step will be 
to include the terms $\sim \Delta_{21}/\Delta_{31}$. 
The numerical  work is in progress. 

\subsection
{Studying Neutrino Oscillation parameters with pure $\nu_e$ beams\\
A. Bandyopadhyay, S. Goswami, P. Mehta, S. Uma Sankar}
Radioactive ions accelerated to high energy decay through
beta process and give pure $\nu_e\bar/{\nue}$ beams which 
are called beta-beams in the literature \cite{beta}. 
The advantages in these are \\
$\bullet$ single neutrino flavour \\
$\bullet$ well known energy spectrum and intensity \\
$\bullet$ strong collimation \\
We studied the possibility of using 
low energy beta beams to probe solar neutrino oscillation parameters.
For $\Delta m^2 =7.2\times 10^{-5}$ eV$^2$ 
maximum oscillation for 50 MeV $\nu_e$ at 732 km 
(CERN to Gran Sasso) 
%\vskip .5cm
%\begin{figure}[htb]
%\epsfxsize=6cm
%\centerline{\epsfbox{pvslbe.eps}}
%\caption{Plot of electron neutrino survival probability $P_{ee}$ 
%vs L/E. We show the values of $L/E$ for which there is a  minima 
%in $P_{ee}$ i.e a maximum of the oscillatory term. 
%}
%\end{figure}
%\vskip -0.5cm
However event rates for such low energy beta beams are very low 
at such distances and therefore the conclusion is 
low energy beta beams are not suitable for probing oscillation 
parameters relevant for the solar neutrino problem. 

%$\bullet$ Is it possible to use the Borexino Detector ? 
%
%$\bullet$ JHF to KamLAND ?
%
%$\bullet$ numerical study to be done

\subsection
{Discussion on INO \\ 
S. Chattopadhyaya,
A. Dighe, D. Indumathi, S. Goswami, D. Majumdar, P. Mehta, N. Mondal,
G. Rajasekaran, A. Raychaudhury, D.P. Roy, P. Roy, A. Samanta, 
S. Uma Sankar}

Following short presentations were given

\begin{enumerate}
\item[1)]ICAL and some present/near future experiments
--  {{S. Goswami}}
\item[2)]RPC studies for ICAL@INO --
{ {N. K. Mondal}}
\item[3)]Can we study matter effects with ICAL ?
-- { {D. Indumathi}}
\item[4)]Physics reach of NuMI off-axis experiment
-- {{S. Uma Sankar}} 
\item[5)]$\beta$ beams, superbeams and neutrino factory beams
-- {{P. Mehta}}
\item[6)]Relationship between neutrino and muon energies in QE and Single pion processes\\
-- {{D. P. Roy}}
\end{enumerate}

\subsection
{CPT violation in neutrinos at Longbaseline
 Experiments \\
Amol Dighe and Poonam Mehta}
The interactions of neutrinos may involve a CPT-odd
term of the form ${{\bar{\nu}_L^\alpha
b_{\alpha \beta}^\mu \gamma_\mu \nu_L^{\beta}}}$,
where $\alpha$ and $\beta$ are flavour indices,
and the elements $b_{\alpha \beta}^\mu$ are in general
complex\cite{barger}.
This term modifies the neurino effective Hamiltonian,
which becomes (in the ultra-relativistic limit)
\begin{equation}
H_{\alpha \beta} = \frac{(m m^\dagger)_{\alpha \beta}}{2 E} +
b^0_{\alpha \beta}~~.
\end{equation}
Here $m$ is the neutrino mass matrix in the flavour basis.
CPT violation arises from the fact that $b^0_{\alpha \beta}$
changes sign when going from neutrinos to antineutrinos.

The special case of two neutrino mixing, where in addition
$m m^\dagger$ and $b^0$ are both diagonalized by the same
unitary matrices, has already been studied.
It has been shown that at future neutrino factories or
at detectors with muon charge identification capabilities,
one can constrain the difference
in $b^0$ eigenvalues to $\delta b \lsim 10^{-23}$ GeV
\cite{barger,gandhi}.
The constraints from reactor and solar neutrino
experiments have also been estimated \cite{bahcall}.

The general case, where the structure of $b^0$ is indepenent
of the structure of $m m^\dagger$, needs to be studied.
This involves taking into account the extra mixing angle
as well as an additional complex phase between the diagonalizing
unitary matrices.
In that general framework, one should first calculate the
current limits on the elements of $b^0_{\alpha \beta}$,
if any, from the solar, atmospheric and reactor data available.
Then one can proceed to compare the extent of these effects
at long baseline experiments and neurino factories,
and devise strategies for disentangling these
from matter effects and CP violating effects at these detectors.

\subsection
{Neutrino masses with Abelian Horizontal Symmetries \\
S. Goswami, D. Indumathi, Y. Shadmi, Y. Nir}
The mass hierarchy in the charged lepton and quark sectors
and the small quark mixings call for an explanation. One of the
most attractive frameworks that provide a natural explanation is
that of the Froggatt-Nielsen (FN) mechanism. One assumes an Abelian
horizontal symmetry that is spontaneously broken at a scale somewhat
lower than some high flavour scale $M_F$. We studied the application
of such models to the neutrino sector, with the hope of explaining the
observed neutrino mass and mixing parameters. If the neutrinos are
Majorana particles then there is an additional scale, $M_L$, which is
the
scale of lepton number violation. The existence of this high scale
explains
the smallness of neutrino masses through the see-saw mechanism. We aim
to explore if the existence of this  additional scale, on top of the
Froggatt-Nielsen scale, can explain the special flavor features that
are
observed in the neutrino sector. We studied some specific examples in a
simplified two neutrino framework during WHEPP.
This work was further continued in \cite{yy}. It was found that
the presence of this scale in the framework of the FN mechanism
can give rise to flavour parameters that have a different
hierarchy than the charged fermions. Moreover, unique features,
such as inverted hierarchy or pseudo-Dirac state, can appear in
the neutrino sector.
We continue to correspond and exchange ideas related to this topic.

\subsection 
{Neutrino masses and mixings in warped extra dimensions \\
K.~Agashe, P.~Das, A.~Dighe, P.~Mehta and P.~Roy}
Consider the Randall-Sundrum (RS1) model \cite{rs1}
which is
a compact slice of AdS$_5$,
\begin{eqnarray}
ds^2 & = & e^{-2k |\theta| r_c} \eta^{\mu \nu} dx_{\mu} dx_{\nu} + r_c^2 d
\theta^2, \; - \pi \leq \theta \leq \pi,
\label{metric}
\end{eqnarray}
where the extra-dimensional interval is realized as an orbifolded circle of
radius $r_c$. The two orbifold fixed points, $\theta = 0, \pi$, correspond
to the ``Planck'' and ``TeV'' branes respectively. In
warped spacetimes the relationship between
5D mass scales and the mass scales in the effective 4D description
depends on the
location in the extra dimension through the
warp factor, $e^{-k |\theta| r_c}$. This allows large 4D mass hierarchies to
naturally arise without large hierarchies in the defining 5D theory, whose
mass parameters are taken to be of the order of  the observed
Planck scale, $M_{ Pl } \sim 10^{18}$ GeV.
For example, the 4D massless graviton
mode is localized near the Planck brane
while Higgs physics is taken to be
localized on the TeV brane. In the 4D effective theory one then
has $M_{\rm weak} \sim M_{ Pl } e^{-k \pi r_c}$,
and
%
%\begin{equation}
%{\rm Weak ~Scale} \sim M_{ Pl } e^{-k \pi r_c} .
%\end{equation}
%
a modestly large radius,
$k \pi r_c \sim \log \left( M_{ Pl } / \hbox{TeV} \right)
\sim 30$, can then accommodate a TeV-size weak scale.
%Kaluza-Klein (KK) graviton resonances have
%$\sim k e^{ - k \pi r_c }$, i.e., TeV-scale masses since their wave functions
%are also localized near the IR brane.
%
%In the original RS1 model, it was assumed that the entire SM (i.e., including
%gauge and fermion fields) is on TeV brane. Thus, effective UV cut-off
%for gauge and fermion fields
%and hence the scale suppressing higher-dimensional operators is
%$\sim$ TeV, i.e., same as for Higgs. However,
%bounds from electroweak precision data
%on this cut-off are $\sim 5-10$ TeV, whereas those from flavor changing neutral 
%currents
%(FCNC's)
%(for example, $K - \bar{K}$ mixing)
%are $\sim 1000$ TeV. Thus, to stabilize the electroweak scale requires
%fine-tuning,
%i.e., even though RS1 explains the big hierarchy between
%Planck and electroweak scale, it has a ``little'' hierarchy problem.
%
We consider a scenario where the SM gauge and fermion fields are
allowed to be in the bulk,
% whereas the Higgs is confined to the TeV brane. We
and show that this can natutrally give the observed
pattern of neutrino masses and mixings.
%A solution to this problem is to move SM gauge and fermion fields into the bulk.
 
The profile in the fifth dimension of the wavefunction of the massless
chiral mode of a $5D$ fermion (identified with the SM fermion)
is controlled by the $c$-parameter \cite{neubert,gp}.
%In the warped scenario,
For $c>1/2$ ($c<1/2$) the zero mode is localized near the Planck
(TeV) brane, whereas for $c = 1/2$, the wave function is  flat.
We choose $c > 1/2$ for light fermions so that the
effective ultraviolet cutoff is much greater than TeV and the
FCNC's are suppressed \cite{gp,huber1}.
Also this naturally results in  small $4D$ Yukawa couplings
to the Higgs on the TeV brane without any hierarchies in the
fundamental $5D$ Yukawa coupling\cite{gp,huber1}.

We introduce $3$ right-handed neutrinos ($\nu_R$'s) in the bulk,
which have identical $c$ values.
% so that their profiles are identical.
The lepton number is taken to be broken only on the Planck
brane, such that these $\nu_R$'s have Majorana mass terms
localized on the Planck brane \cite{huber2}, and hence of the same order
as the Planck scale.
We choose the $\nu_R$ basis that diagonalizes the Majorana mass matrix,
so that
\begin{eqnarray}
M_R & \propto &
\left(
\begin{array}{ccc}
{\cal O}(1) & 0 & 0 \\
0 & {\cal O}(1) & 0 \\
0 & 0 & {\cal O}(1)
\end{array}
\right)  ~~.
\label{Majorana}
\end{eqnarray}

The $\nu_L$s and $\nu_R$'s give rise to a Dirac mass term
\begin{eqnarray}
\left( m_D \right)_{ ij }
& \sim & \left( \lambda_{ 5D } \right)_{ ij } v
\left( \frac{ \hbox{TeV} }{ M_{ Pl } } \right)^{ c_{i  L } + c_{ j R } - 1 }
\label{dirac-term}
\end{eqnarray}
where $\lambda_{ 5D }$ is the $5D$ Yukawa coupling to Higgs
(localized on the TeV brane) and the
last factor comes from the profiles of the $\nu_L$ and
$\nu_R$ zero-modes.
We assume that the matrix $\lambda_{ 5D }$ is anarchic (i.e.,
all its entries are of the same order) and
$c_{\nu_{\mu L}} =  c_{\nu_{\tau L}} < c_{\nu_{eL}}$.
%that $c$'s are same
%for $\nu_L^{ \mu, \tau }$, but larger for $\nu_L^e$,
%whereas $c$'s are same for $\nu_R$'s.
This gives
\begin{eqnarray}
m_D & \propto &
\left(
\begin{array}{ccc}
{\cal O} (\epsilon) & {\cal O}(\epsilon) & {\cal O}(\epsilon) \\
{\cal O}(1) &{\cal O} (1) & {\cal O}(1) \\
{\cal O}(1) & {\cal O}(1) & {\cal O}(1)
\end{array}
\right) ~~,
\label{Dirac}
\end{eqnarray}
where
% (and also in what follows) it is understood that
%there can be an $O(1)$ fluctuation in each entry.
$\epsilon \sim \left( \hbox{TeV} /
M_{Pl} \right)^{ c_{ \nu_{e L}} - c_{ \nu_{ \mu L } } }\ll 1$.
%is due to $c_{ \nu_L^e } > c_{ \nu^{ \mu }_L }$, i.e.,
%wavefunction of $\nu^e_L$ at the TeV brane
%is suppressed compared to that of $\nu^{ \mu, \;
%\tau }_L$.
 
%We assume that lepton number is broken only on the Planck brane
%so that there are {\em anarchic} Majorana mass terms localized
%on Planck brane
%(so that these mass
%terms are of Planckian size) for right-handed neutrinos \cite{huber2}.
%However, the Majorana mass for {\em zero}-mode $\nu_R$'s
%can be a bit smaller than Planck scale (as required to explain
%neutrino data) due to profile of zero-modes of $\nu_R$'s \cite{huber2}
%and the mass matrix can be {\em diagonalized} by a rotation of the
%$\nu_R$ zero-modes:
%
%\begin{eqnarray}
%M^{ diagonal }_R & \propto &
%\left(
%\begin{array}{ccc}
%1 & 0 & 0 \\
%0 & 1 & 0 \\
%0 & 0 & 1
%\end{array}
%\right)
%\label{Majorana}
%\end{eqnarray}
%
%Recall that $c$'s are same for $\nu_R$'s so that the three eigenvalues
%of the above matrix are of same size.
 
Using Eqs. (\ref{Majorana}) and (\ref{Dirac}),
we obtain the following see-saw formula
for Majorana mass for light (mostly left-handed) neutrinos:
\begin{equation}
m_{ \nu }  = m_D M_R^{ -1 } m_D^T \propto
\left(
\begin{array}{ccc}
{\cal O}(\epsilon^2) & {\cal O}(\epsilon) & {\cal O}(\epsilon) \\
{\cal O}(\epsilon) &{\cal O}( 1) &{\cal O}( 1) \\
{\cal O}(\epsilon) & {\cal O}(1) & {\cal O}(1)
\end{array}
\right) ~~.
\end{equation}
 
%\begin{eqnarray}
%m_{ \nu } & = & m_D M_R^{ -1 } m_D^T \nonumber \\
%& \propto &
%\left(
%\begin{array}{ccc}
%\epsilon^2 & \epsilon & \epsilon \\
%\epsilon & 1 & 1 \\
%\epsilon & 1 & 1
%\end{array}
%\right)
%\end{eqnarray}
The larger of the two eigenvalues of the (2--3) submatrix of
$m_\nu$ will naturally be ${\cal O}(1)$.
If the smaller eigenvalues is ${\cal O}(\epsilon)$,
then $m_\nu$ can be diagonalized by
$U \equiv R(\theta_{12}) R(\theta_{13}) R(\theta_{23})$
where
%
%We can show that this gives
\begin{equation}
\theta_{ 23 }  \sim  {\cal O}(1) \quad ,
\theta_{ 13 }  \sim  {\cal O}( \epsilon ) \quad ,
\theta_{ 12 }  \sim  {\cal O}(1) \quad ,
%\theta_{ 12 }  \sim  \arctan \left(
%\frac{ {\cal O}(1) }{ \epsilon } \right) ~,~
\end{equation}
%
%
%
%\begin{eqnarray}
%\theta_{ 23 } & \sim & O(1) \nonumber \\
%\theta_{ 12 } & \sim & \arctan \left( \frac{ O(1) }{ \epsilon } \right)
%\nonumber \\
%\theta_{ 13 } & \sim & O( \epsilon )
%\end{eqnarray}
%
so that we obtain two large
mixing angles and one small mixing angle.
Hence, we get a good fit to data for part of the parameter space
(as shown in \cite{huber3} for a different neutrino set-up in RS1).
The requirement of ${\cal O}(\epsilon)$ eigenvalue of the (2--3)
submatrix of $m_\nu$ may be viewed as a mild fine tuning. Ways of
getting around this would be looked for in the continuation of
this project.

\subsection
{Radiative Magnification of mixing for quasi degenerate neutrinos: Effects
of threshold corrections \\
R.N. Mohapatra, M.K. Parida, G. Rajasekaran}
In \cite{rpr} it was demonstrated that unification of neutrino mixings with 
quark mixings takes place at high scales, but experimental data at low energies 
are explained due to radiative magnification in the neutrino sector provided allthree neutrinos are quasidegenerate and possess the same CP. 
They obtianed $\Delta m^2_{21} \simeq (1.2 - 6.0) \times 10^{-4} $ eV$^2$ 
which is on the higher side of the solar neutrino data. In this work 
, completed during the WHEPP8 workshop they have estimated the 
threshold corrections. 

Threshold effects on the mass basis is defined as 
\beq
m_{ij} = m_i \delta_{ij} + m_i I_{ij} + m_j I_{ji}
\eeq 
A real transformation matrix U is used to express loop factors in the mass 
basis in terms of those in the flavour basis 
\beq
I_{ij} = \sum_{\alpha,\beta} U_{\alpha i} U_{\beta j} I_{\alpha \beta}
\eeq 
The following model independent relations are derived using 
$\theta_{atm} = \pi/4$, $\theta_{CHOOZ} =0$ under the assumption of minimal 
flavour violation with $I_{\alpha \beta} = diag{I_e, I_\mu,I_\tau}$
\be
(\Delta m^2_{\odot})_{\rm th} \simeq 4 m^2 \cos2\theta_{\odot} [-I_e + \frac{1}{2}(I_\mu 
+ I_\tau)]
\ee 
\be
(\Delta m^2_{atm})_{\rm th} \simeq 4 m^2 \sin^2\theta_{\odot} [-I_e + \frac{1}{2}
(I_\mu + I_\tau)]
\ee
\be 
(\Delta m^2_{\odot})_{\rm th} \simeq (\cot^2\theta_{\odot} -1)(\Delta m^2_{atm})_{\rm th}
\ee 
With allowed values of $\theta_{\odot}$ = 18 -40 $^{\deg}$, 
one gets $(\Delta m^2_{\odot})_{th} = (8.4 - 0.4)(\Delta m^2_{atm})_{th} $
demonstrating that both solar and atmospheric data cannot be explained purely by threshold effects. The RG effects and threshold effects both play 
important role in explaining the data.

Using sneutrino/chargino loops they evaluate threshold effects in MSSM 
for $M_{e} \leq 300$ GeV and $M_{\tilde \mu}, M_{\tilde\tau} \simeq (1.2 -1.6) 
M_{\tilde e}$ yielding 
$\Delta m^2_{21} \simeq - 2.7 \times 10^{-5}$ eV$^2$ to $ -5.7 \times 10^{-4} $eV$^2 $ \\
$\Delta m^2_{32} \simeq - 8 \times 10^{-6}$ eV$^2$ to - $ -3.8 \times 10^{-4} $eV$^2 $ \\
Such threshold effects added to the RG corrected values in \cite{rpr} 
gives good agreement with data. 
Further work is in progress. 

\subsection{Supersymmetric Triplet Higgs Model for $M^{\nu}$
%, lepton flavour violation
 and high energy collider test \\
E.J. Chun, J. Forshaw, R. Godbole, Y.Y. Keum, S. Rindani, H. Vaidya}

In scenarios with SUSY and SU(5) gauge unification neutrino masses can arise 
from light Triplet scalars. Such models can be tested in high energy colliders 
from correlated search for bilepton and leptoquarks. 
It was propsed to study the feasibility of such signals through  
numerical calculations.  
  
\subsection{Leptogenesis} 
Some talks and disucssions led by E.J. Chun were held 
on the topic of leptogeneis. Leptogenesis in models with 
two right handed neutrinos and  in the context of type I $\oplus$ type-II 
seesaw scenarios was discussed. 
Standard leptogenesis vis-a-vis soft leptogenesis was studied. 
Some ideas for non-thermal leptogeneis e.g. Inflaton decays 
to right handed neutrino, Domain wall expansion, possible roles of 
inflatino were explored. 

\section{Astroparticle Physics}
The theme of the Astroparticle Physics discussion at WHEPP-8 was particle
physics implications for cosmology.  With the recent interest in branes
and the exciting results from the Wilkinson Microwave Anisotropy Project
(WMAP) the two plenary talks were on {\it Brane World Cosmology} and
on {\it Particle Physics Implications of WMAP Measurements} by Cliff Burgess
and Urjit Yajnik respectively.  The Working Group discussions on particle
physics and cosmology were led by Manoj Kaplinghat, Ajit Srivastava and
Cliff Burgess who discussed {\it Current and Future Measures of Absolute
Neutrino Mass from Cosmology, Topological Defects} and {\it Brane Cosmology}
respectively.  In addition there was a seminar on {\it B-L Cosmic Strings
and Baryogenesis} by Narendra Sahu.
Subsequent discussions led to the following topics being
proposed as areas of study: \\
%\item
$\bullet$ The impact of varying properties of neutrinos
such as mass, mean free path, etc.
on
cosmic microwave background radiation fluctuations
\\ 
%\item
$\bullet$ The effect of cosmic strings on primordial nucleosynthesis
\\ 
%\item
$\bullet$ Constraining primordial black hole abundances from the CMBR
\\ 
%\item
$\bullet$ Inflation and reheating in the context of branes
\\ 
%\item
$\bullet$ The phenomenology of SUSY ADD and intermediate scale string theories
\\ 
%\item
$\bullet$ Exploring the Randall-Sundrum model in higher co-dimensions
\\ 
%\item
$\bullet$ Dark energy and SUSY ADD

After further discussions the efforts of the group were focused primarily
on understanding how
CMBR fluctuations are affected by various particle physics inputs, such as
neutrino properties, the existence of cosmic strings, etc.,
and on explaining dark energy in the context of SUSY ADD models.

\subsection{Particle Physics and CMBR \\
A. Dighe, M. Kaplinghat, Y.Y. Keum, B. Layek, N. Mahajan,
S. More, R. Rangarajan, N. Sahu, A.M. Srivastava
}

Observed fluctuations in the CMBR are a function of the initial fluctuations
in the energy density in various species at the time of decoupling,
and other cosmological factors such as the expansion rate of the universe,
the reionisation of the universe, etc. that affect the evolution of the
photons as the universe evolves from decoupling till today \cite{cmbr}.
Changing the
initial fluctuation spectrum, introducing newer fluctuations after decoupling
or amending the conditions in the universe as the photons propogate after
decoupling can affect the observed CMBR fluctuations today.  Such
studies require massive numerical investigation.  Fortunately, there
exists a publicly available code in Fortran/C called CMBFAST
developed by U. Seljak and M. Zaldarriaga that evolves
fluctuations in photons from decoupling till today while incorporating
various inputs into the evolution.
Effects associated
with quintessence, a 5 dimensional world, interacting dark matter and
gravitational lensing of the CMBR can also be incorporated.
It was decided to first understand the logic
of this code and to then modify it to study various issues.
Below we first paraphrase
a tutorial on CMBFAST
presented by Manoj Kaplinghat.
We then present the list of issues that were suggested for study and the
modifications in the code that these entailed.  Lastly we discuss the
modifications that could actually be carried out during WHEPP-8.
\\ \\ 
%\vskip 0.5in
{\bf CMBFAST}
\\ 
The main program of CMBFAST is cmbflat (in cmbflat.F) or cmbopen
(in cmbopen.F) for a flat or an open/closed
universe respectively.
(For an extreme closed universe see the documentation.)
These programs call subroutine fderivs that is responsible for
evolving perturbations in the energy densities and pressure of baryons,
non-baryonic cold dark matter, hot dark matter
photons and neutrinos, and the metric perturbations.  The various input parameters
for the
program are provided in a file called cmb.par.
These include, for
example, the energy densities today of the various species listed above,
the vacuum energy density today,
the Hubble parameter today, the nature
of initial perturbation in these energy
densities, the number of relativistic and non-relativistic
neutrinos, reionisation parameters, etc.
(See
/EXAMPLES/IARGC/README and /DOC/README\_IAGRC,
and /EXAMPLES/IARGC/cmb.par for a sample input parameters file.)
\footnote{The README files
contain fewer parameters than the sample cmb.par, and the latter
contains fewer parameters than the total set of input parameters.
The explanation for the parameters
is in the documentation.  Parameters that are not specified in cmb.par
take the default values specified in subroutine setpar in params.f.}
The entire
code is
run by driver.F.  The code is compatible with parallel programming.
The equations underlying the code may be obtained from
Ma and Bertschinger, Ap. J. 455 (1995) 7 (astro-ph/9506072).  The notation
used in the code is similar to that in the above paper.

The CMBFAST package may be downloaded from http://www.cmbfast.org .
To start using this package, after unzipping and untarring it, type
./configure $-\,-$with-cobe=yes $-\,-$with-iargc=yes, which generates
a Makefile, and then type
make to compile and create an executable file, cmb.
\footnote{On some platforms additional lines have to be added to
the Makefile generated by configure to explicitly include commands
to compile cmbflat.F, cmbopen.F, driver.F, subroutines.F,
jlgen.F, ujlfen.F and jlens.F to prevent errors while executing the Makefile.
Also on some platforms the program recfast.f has to be amended to avoid syntax
errors associated with the line continuation character placed away from
column 6.
The RAM address extension option may also have to be invoked by
replacing f77 in the Makefile
by f77 -q64 on IBMSP or by f77 -oExt on Sun Workstations, etc.
to avoid errors due to insufficient memory while running CMBFAST.
}
(Additional information on configuration options is available by typing
./configure $-\,-$help .)

The code requires tables of spherical
Bessel functions.  These
are generated by typing ./jlgen, ./ujlgen and ./jlens .  The input parameters
maximum $l$ and $keta$ were taken to be 1500 and 3000 by us and the output
was stored in jl.dat, ujl.dat and jlens.dat respectively (the output filenames
are an input in cmb.par).
$keta$ is $k\eta_0$ where $k$ is the wave number for the Fourier mode
being evolved and $\eta_0$ is the present conformal time. (Conformal
time is related to the usual co-ordinate time $t$ by $d\eta=dt/a$ where
$a$ is the scale factor.)
Maximum $l$ and $keta$ should be larger than the corresponding values,
$akmax0$ and $lmo$, entered in cmb.par.
Note that $akmax0$ should be larger
than about $2 lmo$ to ensure accurate integration over $k$.

Now copy the cmb.par file from
/EXAMPLES/IARGC/cmb.par to the directory with the executable file
cmb and amend the file as required.
Run the code by typing ./cmb cmb.par .\footnote
{If you get a Segmentation fault try changing nstep0=7000 to nstep0=2400
in cmbfast.inc and do make again.  This is probably related to insufficient
RAM.  See the previous footnote for RAM address extension.}
The output $C_l$ coefficients for the temperature fluctuations
are stored in cl\_unlensed.d and cl\_lensed.d (filenames are set through
variables in cmb.par), which can then be plotted.
The documentation contains information about other output.
\\

We now list below the particle physics issues that the group discussed
and instructions on how
to change the code to calculate their impact on the CMBR.
The suggested modifications to the code may not be complete.
\\ \\
%\item
$\bullet${\bf Neutrino masses}
\\ \\
Neutrinos affect the evolution of the background
photons in the universe in several ways.
Firstly, they contribute to the energy density of the universe and
hence the expansion rate.  Secondly, free streaming neutrinos suppress
anisotropies in matter and in photons for scales less than the mean free
path of the neutrinos.  Thirdly, the latter effect also affects the lensing
of the CMBR by matter in the universe.  Fourthly, fluctuations in the energy
density in neutrinos at decoupling can
affect the fluctuations in the energy density in photons.
(See, for example, Refs. \cite{BS,chackoetal} for discussions of the effects
of neutrinos on CMBR fluctuations.)
 
The energy density in neutrinos is a function of the mass.
Neutrinos with masses
$10^{-4} {\rm eV}\leq m_\nu \leq 1 {\rm MeV}$
decouple when they are relativistic
but are non-relativistic today and their
energy density
today is given by \cite{kolbturner}
\begin{equation}
\Omega_{\nu nr} h_0^2={m_\nu\over 93.5 {\rm eV}}
{10.75\over g_*}\times N_{\nu nr}
\end{equation}
where $N_{\nu nr}$ is the number of species of non-relativistic neutrinos
and $g_*$ is the number of relativistic degrees of freedom when
massive neutrinos
decouple.
Thus to study the
effect of the neutrino mass on the CMBR,
the parameters to be entered into the code are
$\Omega_{\nu nr}$, $N_{\nu nr}$ and $g_*$. These correspond to the variables
$omegan$, $annunr$ and $gsnunr$ in cmb.par.
\footnote{$gsnunr$ does not appear in the sample cmb.par files.  You may
include it in cmb.par, otherwise CMBFAST uses the default value of 10.75
set in params.f.}
The neutrino mass, normalised to the temperature of relativistic neutrinos today,
is calculated
from the input $\Omega_{\nu nr}$ and $N_{\nu nr}$
in cmbflat.F or cmbopen.F
(search for
$amnu$).
Currently the CMBFAST code only
allows all species of non-relativistic neutrinos to have the same mass.
Future work could include modifying the code to allow for neutrinos of differentmasses.
\\ \\
%\item
{\bf $\bullet$ Time varying neutrino masses} \\ \\
Time varying neutrino masses affects the energy density in neutrinos,
perturbations in the energy density, the pressure
and perturbations in the pressure.  To incorporate this
requires modifying the form of the variable $amnu$ in subroutine
initinu1 in subroutines.F .
Certain changes may also be required in subroutines
nu1, nu2, ninu1 and nuder in subroutines.F.
\\ \\
%\item
{\bf$\bullet$  Mean free path of neutrinos}
\\ \\
As mentioned earlier, this affects the spectrum of density perturbations
in photons and matter.  Varying the mean free path of neutrinos can be
achieved by modifying
the right hand side of the evolution equations for perturbations
in neutrinos in fderivs.
(Search for the string `equations of motion' in fderivs.)
%Equations for perturbations in the metric will relate the perturbations
%in neutrinos to those in matter.
\\ \\
%\item
{\bf $\bullet$ Decaying neutrinos}
\\ \\
This affects the evolution of the total energy density, $grho$, and
pressure, $gpres$, in fderivs by changing the contribution of the
neutrinos and the decay products.  For the neutrinos,
this can instead be incorporated by modifying
$rhonu$ and $pnu$ in subroutine ninu1 in subroutines.F.
Changes may also be needed in nuder in subroutines.F.
Decaying neutrinos affects the evolution equation for perturbations in
the neutrinos and their decay products,
which requires modifying the right hand side of the corresponding
equations in fderivs and by modifying subroutine nu2 in subroutines.F.
(Search for the string `equations of motion' in fderivs.)
Decaying neutrinos has been considered earlier in,
for example, Ref. \cite{manoj}.
\\ \\
%\item
{\bf $\bullet$  Cosmic strings}
\\ \\
Cosmic strings seed fluctuations in the photons continuously as they move
through the background.  This seed fluctuation may be modelled and inserted
in subroutines fderivs and
finitial where the string `Add a seed if desired' appears.
\\ \\
%\item
{\bf $\bullet$ Large extra dimensions}
\\ \\
The existence of more than 3 large spatial dimensions affects the expansion
rate of the universe.  This requires changes in the variables $adot$ and
$adotdot$ in the subroutines fderivs, fderivst and finitthermo.
These have already been incorporated in the code and the necessary
changes in the code have been marked with `DIM'
in the subroutines.  CMBR observations in the context of large extra dimensions
have been studied in, for example, Ref. \cite{dim}.
\\ \\
%\item
{\bf $\bullet$ Interacting dark matter}
\\ \\
This can affect the evolution of the energy density and pressure of the dark
matter and necessary changes must be made self-consistently throughout
the code (e.g. in $grho$ and $gpres$ in fderivs).
This also affects the evolution of the perturbations
in dark matter
and can be incorporated by amending the right hand side for the evolution
equations for dark matter perturbations in fderivs.
(Search for the string `equations of motion' in fderivs.)
The effect of interactions between dark matter and baryons on the CMBR
has been studied in Ref. \cite{chen}.
\\ \\
%\item
{\bf$\bullet$  Decaying light gravitinos}
\\ \\
The effect of this would depend on when the light gravitinos decay and to what
they decay.  To incorporate decaying gravitinos they would
first have to be included in $grho$ and $gpres$ in fderivs
and necessary changes would have to be made to the contribution of the
daughter particles in $grho$ and $gpres$.
To include perturbations,
equations for the evolution of perturbations in the energy density
and pressure of gravitinos would have to be included in fderivs and
changes would also have to be made on the right
hand side of the evolution equations for the daughter particles.
\\ \\
%\item
{\bf $\bullet$  Time varying fundamental parameters}
\\ \\
Any time variation in the fine-structure constant
alters the ionization history of the universe and therefore
changes the pattern of cosmic microwave background fluctuations.
The CMBR fluctuations can also be affected by a time varying
gravitational constant.  A time varying electron mass has a similar effect
as a time varying $\alpha$ in that it also affects the ionisation history of theuniverse.
Time varying $\alpha$, $G_N$ and electron mass have been considered in the past
in, for example, Refs. \cite{alpha,zz,yooscherrer} respectively.
There was no further discussion of these issues at WHEPP-8.
Regarding a time varying gravitational constant, note
that energy densities in CMBFAST are entered as $grho$ which is the
energy density times $(8\pi/3) G_N a^2$, where $G_N$ is Newton's gravitational
constant
(see fderivs).  The pressure, $gpres$, and
perturbations in the energy density and pressure, $dgrho$
and $dgpres$, are defined similarly.
\\

%\vskip 0.5in
Much of the effort at WHEPP-8 was devoted to identifying where modifications
would be needed in the code to incorporate various interesting particle
physics ideas.  These have been listed above.  In the remaining time available,
time varying neutrino masses were studied with a trial parametrisation of the
neutrino mass as below:
\begin{equation}
m(t)=\bar m \, {\rm exp}[-\left(a/a_{nr}\right)^{{1\over n}}]\, ,
\end{equation}
where $a_{nr}$ is the scale factor when the neutrino becomes non-relativistic.
The variation of the CMBR fluctuations for constant mass and $n=10$ are
given in Fig. 1.  However not all changes may have been made consistently
in CMBFAST.
\begin{figure}[htbp]
\epsfxsize=8cm
\centerline{\epsfbox{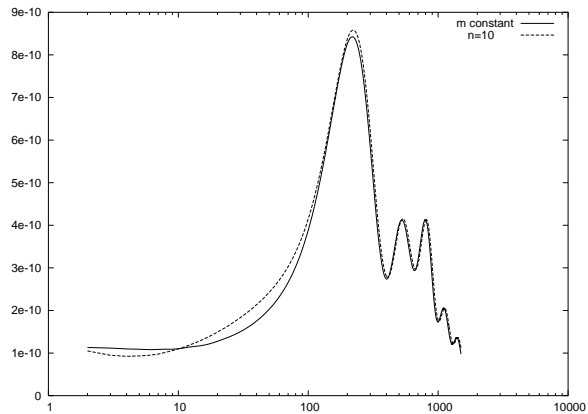}}
\caption{$C_l$ vs $l$ for constant mass and $n=10$
}
\label{fig:food}
\end{figure}
 
The code was also modified to incorporate one massive neutrino species
decaying to
massless neutrinos.
At WHEPP-8 we were able to modify the code to alter the evolution equations
for the decaying particle but not the daughter particles.  One hopes that
this will be addressed in the future.  If the code is modified to include
massive neutrinos of different masses, one could also consider scenarios
where one massive neutrino decays to a lighter massive neutrino.
 
Some effort was made to incorporate cosmic string seeded perturbations
in the CMBR but some issues regarding the nature of the perturbation needed to
be resolved.

\subsection{Brane Cosmology\\
C. Burgess, N. Gupta, Y.Y. Keum, P. Ramadevi, U. Yajnik}

At WHEPP-8 there were some prelimnary discussions on reheating in the context
of branes.  The remaining discussions were on Supersymmetric ADD 
and Dark Energy, as given below.
\\ \\
{\bf $\bullet$ Supersymmetric ADD and Dark Energy}
\\ \\
There have been proposals to explain the origin of the
Dark Energy in a natural way using extra dimensions and 
supersymmetry. The scheme pursued in  \cite{Aghababaie:2003wz}
was reviewed. Its proposal consists of having the observed 
Universe as a brane in 6 dimensional space-time. There are
also other similar branes and the theory to begin with
is supersymmetric. The interest in 6 dimensions is justified from
the ADD \cite{ADD} observation that the current experimental
limits permit a brane world compactification scale $r_c$ as 
large as 2 mm, which corresponds to 
\begin{equation}
\frac{1}{r_c} \sim M_c \sim \frac{M_{EW}^2}{M_{Pl}}\sim 10^{-4}eV
\end{equation}
Supersymmetry ensures a natural value for  the vacuum energy 
to be  zero. The \emph{small} vacuum energy can then be obtained 
as follows. 

The $N=1$ supersymmetry of 6 dimensions manifests 
as $N=2$ supersymmetry in 4 dimensions. The presence of the brane 
breaks translation invariance and hence supersymmetry.
However it has been shown \cite{PolHug86}
that the breaking of translation invariance leads to the breaking
of precisely one of the two supersymmtries on the brane.
Which supersymmetry breaks is decided by the winding number
of the brane treated as a vortex solution in 6 dimensions.
If there are several branes present it can be arranged that
another brane has the complementary supersymmetry broken due
to opposite winding number. As a result, the complete supersymmetry
of the 6 dimensions is broken. However each of the branes possesses 
an  approximate supersymmetry, broken to a small extent by 
the breaking caused on the other brane.
On phenomenological grounds we expect the supersymmetry on our
brane to be broken at the electroweak scale $M_{EW}$. It has
been argued that in this case the splitting of the gravity
supermultiplet in the bulk would be of the order 
$\Delta M_{SUSY}\sim (M_{EW}^2)/M_{Pl}$.

One can now compute the conserved vacuum energy in the following scheme.
\begin{eqnarray}
\lambda_{eff}&=&\textrm{brane tensions}+\textrm{bulk curvature}
+\textrm{bulk loop corrections} \\ \nonumber
&=& \sum_i T_i + \int d^2y \sqrt{-g}(R + F^2 + ...)
\end{eqnarray}
However the source of the bulk gravity are the branes, so that
the first two terms cancel. Then the bulk loops provide
\begin{equation}
\lambda_{eff} \approx (\Delta M_{SUSY-bulk})^4 \sim 
\left(\frac{M_{EW}^2}{M_{Pl}}\right)^4
\end{equation}
being precisely $(10^{-4}eV)^4$ matching the observed vacuum
energy and also within the experimental limits of gravity 
experiments.
 
One important issue faced by the scenario is stabilising the
distance scale between the branes so that the variation of the
vacuum energy is slow enough to not contradict the approximately
constant value of current cosmology.

There are further contributions to the $\lambda_{eff}$ from subleading 
terms. The problem is a systematic assessment of these contributions.
In \cite{Aghababaie:2003wz} these contributions were computed
in Salam-Sezgin\cite{SalSez} 6D Supergravity model. It was shown
that we can order these contributions in powers of a cutoff 
$\Lambda$.
\begin{equation}
\delta V_{eff} \sim \Lambda^6 + \Lambda^4 + \frac{\Lambda^2}{r_c^2}
\end{equation}
where the first two terms can be shown to be well controlled but
the third needs further verification. It would be interesting to
pursue a calculation within the Supergravity model compactified
on a two-sphere, ${\mathbb R}^4\otimes S^2$, with two branes
of opposite winding numbers present.

\acknowledgements
S.G. would like to acknowledge help from  
K. Agashe, A. Dighe, Y. Nir and P. Mehta   
in preapring the section on Neutrino Physics.  
R.R. would like to thank Manoj Kaplinghat, Surhud More and Urjit Yajnik for
help in preparing the section on Astroparticle Physics.    

%\end{enumerate}

\end{document}